\documentstyle[12pt,psfig]{article}
\begin{document}
\title{\hfill UGI-97-29\\[5mm]
Meson $m_T$-scaling in heavy-ion collisions at SIS energies
\thanks{Work supported by BMBF and GSI Darmstadt.}}
\author{E. L. Bratkovskaya, W. Cassing and U. Mosel\\[5mm]
Institut f\"ur Theoretische Physik,
Universit\"at Giessen\\D-35392 Giessen, Germany }
\date{}
\maketitle

\begin{abstract}
We perform systematical studies on transverse-mass spectra of $\pi^0$,
$\eta$, $\omega$, $\phi$, $K^+$ and $K^-$ mesons for C~+~C, Ni~+~Ni,
Au~+~Au collisions at SIS energies within the HSD transport approach.
We find that the $m_T$-spectra  sensitively reflect the in-medium
properties of the mesons. The 'bare mass' scenario leads to a general
scaling behaviour for the meson $m_T$-spectra when including a mass
shift due to the associate strangeness production threshold for kaons
and antikaons whereas a 'dropping' mass scheme violates this
$m_T$-scaling. The relative slope parameters, however, provide valuable
information about the average meson potentials in the nuclear medium.
\end{abstract}

\vspace*{1cm}


\newpage

The properties of hadrons in dense nuclear matter are of
fundamental interest for our understanding of the strong interaction.
QCD sum rules  \cite{H&L92} and QCD
inspired effective Lagrangian models
\cite{BrownRho,Klingl96,Friman}\cite{RappNPA,Peters,Kaplan,Brown1,Waas}
predict changes of the vector ($\rho$, $\omega$, $\phi$) and
strange  ($K^+, K^-$) mesons with increasing nuclear density, i.e.
a modification of their in-medium dispersion relation. Furthermore, due
to scattering and absorption processes at finite baryon density the
width of the mesons is expected to change considerably, too.

The in-medium properties of vector mesons have been studied
experimentally so far by dilepton measurements at SPS energies for
proton-nucleus and nucleus-nucleus collisions
\cite{CERES,Ullrich,HELIOS}.  The observed enhancement in A + A
reactions compared to p + A collisions in the invariant mass range $0.3
\leq M \leq 0.7$ GeV might be explained by a 'dropping' $\rho$-meson
mass following Brown/Rho scaling \cite{BrownRho} or the sum rule
prediction by Hatsuda and Lee~\cite{H&L92} (cf. Refs.
\cite{Li,Cass95C,Brat97,QM96}).  However, as found in Ref.
\cite{CBRW97}, the SPS dilepton data are also compatible with a
hadronic scenario when using a $\rho$-meson spectral function that
includes the pion modifications in the nuclear medium due to
resonance-hole loops as well as the polarization of the $\rho$-meson
due to resonant $\rho-N$ scattering \cite{RappNPA}.

Since the pioneering work of Kaplan and Nelson \cite{Kaplan} predicted
a kaon condensation in nuclear matter, a lot of theoretical efforts
have been devoted to study the properties of strange mesons at finite
baryon density.  According to Refs.~\cite{Kaplan,Brown1,Waas} antikaons
should feel strong attractive forces in the medium whereas the kaon
potential is expected to be slightly repulsive at finite nuclear
density. Kaon spectra and flow have been measured experimentally at
SIS, AGS and SPS energies (cf. Ref. \cite{QM96}).  The comparison of
transport model results with experimental data at SIS energies
\cite{Schro,KaoS,FOPI} shows that the antikaon spectra can not be
described without a sizeable attractive potential (cf.
\cite{Kolix,CBMTS97km,LiLB97}), whereas the kaon flow observed by the
FOPI collaboration \cite{FOPI} indicates a slightly repulsive kaon
potential (cf. \cite{flow1,BCM97kp}).  Thus there are several
experimental indications for the changes of the meson properties
in-medium. However, it is very desirable to have additional independent
experimental observables and criteria.

Here we propose to study the meson $m_T$-spectra in heavy-ion
collisions, where $m_T$ is the transverse mass $m_T=(p_T^2+m^2)^{1/2}$
and $p_T$ the transverse momentum of a meson with bare mass $m$. The
transverse-mass spectra are a common way to represent experimental
information on particle production in heavy-ion physics \cite{QM96}.
The measurement of $m_T$-spectra is usually associated with studies on
equilibration phenomena of the system due to the trivial exponential
behavior: i.e., if the spectrum $1/m_T^2 d\sigma/d m_T$ is of Boltzmann type
$\sim \exp(-\beta m_T)$, the slope parameter $\beta$ might be related
to the global (inverse) temperature at freeze-out in the absence of flow.

The $m_T$-spectra of $\pi^0$ and $\eta$ mesons in heavy-ion collisions
at SIS energies were measured by the TAPS
collaboration~\cite{TAPSold,TAPS-CC,TAPS-CaNi} and $m_T$-scaling has
been found for both mesons and all systems investigated.  Such a
universal property of the meson spectra at SIS energies had been
predicted already by the Quark-Gluon-String Model calculations in
Ref.~\cite{Toneev} for the Ar~+~Ca system at several energies.  A
systematic analysis of $\pi^0$ and $\eta$ spectra was performed
recently in Ref.~\cite{BCRW97} within the HSD transport approach
\cite{Ehehalt}, where also a dropping mass scheme for $\eta$ mesons has
been studied. It was found that the transport model gives a reasonable
description for the $m_T$-spectra of pions and $\eta$'s  without
incorporating any medium modifications for both mesons, whereas a
dropping mass scheme for $\eta$-mesons violates $m_T$-scaling.

In this letter we examine the $m_T$-spectra of all mesons, that can be
produced at SIS energies, to obtain information about their
in-medium meson properties.  We perform systematical studies for
$m_T$-spectra of $\pi^0$, $\eta$, $\omega$, $\phi$, $K^+$ and $K^-$
mesons for C~+~C, Ni~+~Ni, Au~+~Au collisions at SIS energies within
the same transport approach~\cite{Ehehalt} which has been used
previously for the description of hadronic and dilepton data from SIS
to SPS energies \cite{Cass95C,Brat97,Ehehalt}.  We present calculations
with bare meson masses (i.e. no medium modifications) as well as with
in-medium meson masses, however, discarding an explicit momentum
dependence of the meson selfenergies so far.

In relativistic heavy-ion collisions at SIS energies the nuclei can be
compressed up to about 3 times normal nuclear matter density $\rho_0$.
In the hot compression zone the nucleons are excited to baryonic
resonances which decay by emitting mesons and the produced mesons then
can be absorbed, re-emitted and re-scattered.  Due to the small cross
sections involved we treat the production of vector mesons, $\eta$'s,
kaons and antikaons perturbatively at low energies.  The $\eta$ mesons
are produced in pion-baryon and baryon-baryon collisions according to
the elementary production cross sections from
Ref.~\cite{Wolf90,Vetter}. In the present analysis (as in
\cite{BCRW97}) we assume the $pn \to pn\eta$ cross section to be about
6 times larger than the $pp \to pp\eta$ cross section close to
threshold in line with the new data from the WASA
collaboration~\cite{WASA}.  For the vector mesons we take into account
pion-baryon and baryon-baryon production channels according to the
elementary production cross sections from
Refs.~\cite{SibCM97,Sibirtsev}.  The treatment of $\rho$ and $\omega$
meson production and propagation is described in detail in
Ref.~\cite{WBCM97}.  For kaon and antikaon production we include the
channels $BB \to K^+YN$, \ $\pi B\to K^+Y$, \ $BB \to NN K \bar{K}$, \
$\pi B\to N K\bar{K}$, \ $K^+B\to K^+B$, \ $\bar{K} B\to \bar{K}B$, \
$Y N\to \bar{K} NN$, \ $\pi \pi\to K \bar{K}$ as well as $\pi Y\to
\bar{K}N$ and $\bar{K} N\to \pi Y$ for the antikaon absorption.  All
elementary cross sections are taken as in
Refs.~\cite{CBMTS97km,BCM97kp}. For a more detailed description of the
transport approach as well as the explicit parametrizations of all the
production and absorption channels employed we refer the reader to our
preceeding publications \cite{CBMTS97km,BCM97kp,BCRW97,Ehehalt} and
continue with a direct presentation of our results.

In Fig.~\ref{mtFig1} we show the results of our calculations with
bare meson masses for the transverse-mass spectra of $\pi^0$ (solid
line with solid circles), $\eta$ (dashed line with open circles),
$\omega$ (dot-dashed line with solid 'up' triangles), $\phi$ (dotted
line with solid squares), $K^+$ (short dashed line with open diamonds)
and $K^-$ (dotted line with open 'down' triangles) mesons for C~+~C
collisions at 2.0 A$\cdot$GeV (upper part), Ni~+~Ni at 1.93 A$\cdot$GeV
(middle part) and Au~+~Au at 1.5 A$\cdot$GeV (lower part). For scalar
mesons ($\pi^0,\eta$) and vector mesons ($\omega,\phi$) we use
\begin{equation}
m_T^* = m_T = (p_T^2 + m^2)^{1/2},
\label{mt1}
\end{equation}
whereas for strange mesons ($K^+,K^-$) $m_T^*$ contains the shift
in threshold due to the $\Lambda-N$ mass difference, i.e.:
\begin{equation}
m_T^*(K^+) = \left(p_T^2 + (m_K + m_\Lambda - m_N)^2 \right)^{1/2},
\label{mtKp}
\end{equation}
or associated ($K^+,K^0$) mesons,
\begin{equation}
m_T^*(K^-) = \left(p_T^2 + (2 m_K)^2 \right)^{1/2},
\label{mtKm}
\end{equation}
where $m_K$, $m_\Lambda$ and $m_N$ are the masses of kaon, $\Lambda$
and nucleon, respectively. The contributions of vector mesons $\omega$
and $\phi$ are divided by a factor of 3 due to the 3 different
polarizations of the vector mesons.  As can be seen from
Fig.~\ref{mtFig1} the $\pi^0, \eta, \omega$ and $K^+$ spectra indicate
$m_T$-scaling, whereas the contribution of the $\phi$ meson is
suppressed by a factor $\approx$10, however, it has the same slope as
the other mesons.

The $K^-$ spectra (using (\ref{mtKm})) approximately scale only for
the light system C~+~C; for heavy systems such as Ni~+~Ni (middle part)
and especially Au~+~Au (lower part) the $K^-$ spectra are essentially
below the scaling line due to a stronger absorption by baryons. This
result for $K^-$ is also consistent with our previous analysis
\cite{CBMTS97km} showing that the antikaon yield calculated with a bare
mass significantly underestimates the experimental data
\cite{Schro,KaoS}.  Only when including a sizable attractive antikaon
potential (or dropping antikaon mass with density) we could describe
the data \cite{Schro,KaoS} satisfactorily.

We note that the apparent $m_T$ scaling especially
for C + C at 2 A$\cdot$GeV is not due to thermal and chemical
equilibration as suggested
in Ref. \cite{Cleyman} or in Refs. \cite{Stachel1,Stachel2} for AGS and
SPS energies. Here it merely reflects the fact that the production of a meson
(per degree of freedom), after folding over the baryon-baryon and pion-baryon
collisional distribution in the invariant energy $\sqrt{s}$,
essentially depends on the excess energy available
as suggested by Metag \cite{Metag}.
This notion is also consistent with the mass shifts in (\ref{mtKp}) and
(\ref{mtKm}) for kaons and antikaons, where the shift in threshold due
to the associated strange hadron is taken into account explicitly.  On
the other hand, in chemical equilibrium no shift in the transeverse
mass as in (\ref{mtKp}) and (\ref{mtKm}) should be considered. Our
calculations, furthermore, show a strong anisotropy in the
center-of-mass angular distribution for all mesons similar to that of
pions \cite{Teis,Pelte} which raises severe doubts on the issue of thermal
equilibration, too.

In Fig.~\ref{mtFig2} we show -- again for bare meson masses -- our
calculations for the $m_T$-spectra of $\pi^0, \eta, \omega$ and $K^+$
mesons for C~+~C (upper part) and Au~+~Au (middle part) at
1.0~A$\cdot$GeV.  As seen from  Fig.~\ref{mtFig2} the spectra at
1.0~A$\cdot$GeV show the same scaling behaviour as in
Fig.~\ref{mtFig1}. The lower part of Fig.~\ref{mtFig2} corresponds to
calculations for Au~+~Au at 1.0~A$\cdot$GeV gating on central rapidity
$-0.3\le y \le 0.3$.  The rapidity cut decreases the particle yield,
however, does not 'destroy' the global scaling behaviour. Thus data at
midrapidity provide similar information.

In order to investigate the dynamical origin of the scaling behaviour
found for bare meson masses more closely,
we show in Fig.~\ref{mtFig3} the channel
decomposition for $\eta, \omega, K^+$ and $K^-$ spectra for Au~+~Au at
1.5~A$\cdot$GeV.  The solid lines with solid circles indicate the sum
over all contributions,  the dashed lines with open circles correspond
to the $NN$ production channel, the dotted lines with open squares are
the $\Delta N$ channel, the dot-dashed line with solid 'up' triangles
show the $\pi N$ contribution and the short dashed line with open
diamonds in the lower part of Fig.~\ref{mtFig3} represents the pion-hyperon
($\pi Y$) channel for $K^-$ production.

For $\eta$ production all included channels ($NN, \Delta N, \pi N$)
give practically the same contribution because at 1.5 A$\cdot$GeV  we
are above the pion-baryon and baryon-baryon $\eta$ production
threshold.  For the $\omega$ meson the $\pi N$ channel provides the
dominant yield, the $NN$ and $\Delta N$ contributions are below due to
the large threshold for $\omega$ production in baryon-baryon
collisions.  For $K^+$ production the $\pi N$ and $\Delta N$ channels
are approximately of the same order of magnitude; the
contribution from the $\Delta N$ channel is more important at
1.5~A$\cdot$GeV in comparison to the lower energy of 1.0~A$\cdot$GeV,
where the $\pi N$ channel is dominant (cf. Ref.~\cite{BCM97kp}).  In
line with our previous analysis (cf. Ref.~\cite{CBMTS97km}) the $\pi Y$
channel gives the main contribution for $K^-$ production. This is due
to the fact that in more central collisions the pion density reaches
about $0.15 fm^{-3}$ while the hyperons have almost the same abundancy
as the $K^+$ and $K^0$ mesons.  Thus a substantial amount of hyperons
suffer a quark exchange with pions ($s \to u,d$) when propagating out
of the nuclear medium. On the other hand the same type of process
(i.e. flavor exchange) leads to antikaon absorption on baryons.

Fig.~\ref{mtFig3} demonstrates that the secondary channels $\pi N$ and
$\Delta N$ give the main contribution for meson production at SIS
energies which are very close (or even below) the elementary production
threshold for most of the mesons.  The primary $NN$ collisions become
important only if the incoming energy is substantially above the
elementary production threshold (as in the $\eta$ case).

We now turn to the question of in-medium meson properties by adopting
the scaling hypothesis of Brown and Rho \cite{BrownRho}.
For our calculations we use a linear extrapolation of the meson masses
with baryon density,
\begin{equation}
m^*_M = m_M^0 \left(1 - \alpha_M \frac{\rho_B}{\rho_0}\right),
\label{inmedmass} \end{equation}
with $\alpha_\eta \approx 0.18$ as in Ref. \cite{BCRW97},
$\alpha_\omega \approx 0.18$
\cite{H&L92}, $\alpha_\phi \approx 0.00225$ in line with
\cite{Chung97}, $\alpha_{K^-} \approx 0.24$, $\alpha_{K^+} \approx
-0.06$.  The parameter $\alpha_{K^-} \approx$ 0.24 corresponds to an
attractive potential of about $-120$~MeV roughly in line with
Refs.~\cite{Kaplan,Brown1,Waas} whereas $\alpha_{K^+} \approx -0.06$
leads to a slightly repulsive potential of about $+30$~MeV as in
Ref.~\cite{Brown1}. Since in Eq. (\ref{inmedmass}) we have neglected an
explicit momentum dependence of the meson self energies our following
calculations should be considered as a more qualitative study.

The  symbols in Fig.~\ref{mtFig4} indicate the results from our HSD
calculations for C~+~C at 2.0~A$\cdot$GeV, Ni~+~Ni at 1.93~A$\cdot$GeV
and Au~+~Au at 1.5~A$\cdot$GeV (the assignment is the same as in
Fig.~\ref{mtFig1}).  The solid lines are exponential fits to the HSD
results for orientation.  The pion spectra are shown by the thick
straight lines indicating the general scaling behaviour (cf.
Fig.~\ref{mtFig1}) with the slope parameters 77~MeV for C~+~C at
2.0~A$\cdot$GeV, 82~MeV for Ni~+~Ni at 1.93~A$\cdot$GeV and 83~MeV for
Au~+~Au at 1.5~A$\cdot$GeV.  The 'dropping' of the $\eta, \omega$ and
$K^-$ masses according to Eq.~(\ref{inmedmass}) leads to an enhancement
of the  $m_T$-spectra especially at low $m_T$ due to the shift of the
production thresholds to lower energy.  The enhancement of the $\phi$
yield is almost not seen due to the very small coefficient
$\alpha_\phi$ in (\ref{inmedmass}).  The increasing $K^+$ mass in the
medium (Eq.~(\ref{inmedmass})) leads to a suppression of the $K^+$
yield at low $m_T$ correspondingly.

Thus the in-medium modifications of the mesons according to
Eq.~(\ref{inmedmass}) destroy the $m_T$-scaling picture presented in
Fig.~\ref{mtFig1}. As already shown in Ref.~\cite{BCRW97} the simple
'dropping' $\eta$ mass scheme is not consistent with the $m_T$-scaling
observed by the TAPS collaboration~\cite{TAPSold,TAPS-CC,TAPS-CaNi}.
On the other hand the present experimental data for $K^+$ and $K^-$
\cite{Schro,KaoS,FOPI}  indicate that the in-medium mass scheme for
strange mesons leads to a reasonable agreement with the data. So, one
can expect that the kaon and antikaon $m_T$-spectra should not show a
scaling behavior, i.e.  the slopes should be different from the
pions, especially for $K^-$ mesons.

In summary, our analysis shows that the $m_T$-spectra of mesons
(corrected by production thresholds as in (\ref{mtKp}), (\ref{mtKm}))
are quite sensitive to the in-medium modifications of the mesons.  For
bare meson masses we find a general scaling for $\pi^0, \eta, \omega$
and $K^+$ mesons, whereas $K^-$ and $\phi$ are suppressed, however,
have the same slope. The $m_T$ scaling found within our transport
simulations is not due to thermal and chemical equilibration as
suggested in Ref.  \cite{Cleyman}, but a genuine nonequilibrium effect
which is seen most prominantly for the light system C + C at 2
A$\cdot$GeV. Dynamically it results from the observation \cite{Metag}
that the production probability (per degree of freedom) of the mesons
considered here (except the $\phi$) in heavy-ion reactions 'only'
depends on the excess energy available in hadron-hadron collisions. Any
in-medium modifications thus show up in softer $m_T$-slopes in case of
attractive potentials and higher onsets at $p_T=0$; the opposite holds
for repulsive potentials.  The change in slope relative to pions, that
are usually detected simultaneously with the heavier mesons, is
proportional to the $\alpha$ parameter in Eq.~(\ref{inmedmass}) which
reflects the sign and strength of the meson potentials in-medium.
These relations do not vary much when restricting to midrapidity or
changing the bombarding energy according to the HSD transport
calculations. Thus the $m_T$-spectra should provide valuable
information on the in-medium properties of the heavier mesons once the
pion $m_T$ spectra are measured in the same experiment for reference.

\vspace*{5mm}
The authors acknowledge valuable discussions with N.~Herrmann,
V.~Metag, H.~Oeschler, P.~Senger, A.~Sibirtsev and V.D.~Toneev.

\newpage
\section*{Figure captions}

\begin{figure}[h]
\caption{The calculated inclusive transverse-mass spectra of $\pi^0$
(solid line with solid circles), $\eta$ (dashed line with open
circles), $\omega$ (dot-dashed line with solid 'up' triangles), $\phi$
(dotted line the solid squares), $K^+$ (short dashed line with open
diamonds) and $K^-$ (dotted line with open 'down' triangles) mesons for
C~+~C collisions at 2.0 A$\cdot$GeV (upper part), Ni~+~Ni at 1.93
A$\cdot$GeV (middle part) and Au~+~Au at 1.5 A$\cdot$GeV (lower part).
The calculations are performed for bare meson masses.}
\label{mtFig1}

\caption{Inclusive $m_T$-spectra of $\pi^0, \eta, \omega$ and $K^+$ mesons
for C~+~C (upper part) and Au~+~Au (middle part) at 1.0~A$\cdot$GeV.
The lower part corresponds to Au~+~Au at 1.0~A$\cdot$GeV
for a central rapidity bin $-0.3 \le y \le 0.3$.}
\label{mtFig2}

\caption{The channel decomposition for $\eta, \omega, K^+$ and $K^-$
inclusive $m_T$-spectra for Au~+~Au at 1.5~A$\cdot$GeV.  The calculations
have been performed with bare meson masses. The solid lines with solid
circles indicate the sum over all contributions, the dashed lines with
open circles correspond to the $NN$ production channel, the dotted
lines with open squares are the $\Delta N$ channel, the dot-dashed lines
with solid 'up' triangles show the $\pi N$ contributions while the short
dashed line with open diamonds is the pion-hyperon ($\pi Y$) channel for
$K^-$ production.}
\label{mtFig3}

\caption{The calculated inclusive transverse-mass spectra of $\eta, \omega,
\phi, K^+$ and $K^-$ mesons with in-medium masses according to
Eq.~(\protect\ref{inmedmass}). The symbols indicate the results from the HSD
calculations (the assignment is the same as in Fig.~\protect\ref{mtFig1}).
The solid lines are exponential fits to the HSD results to guide the eye.
The pion yield is shown by straight thick lines.}
\label{mtFig4}
\end{figure}

\newpage
\pagestyle{empty}
\psfig{figure=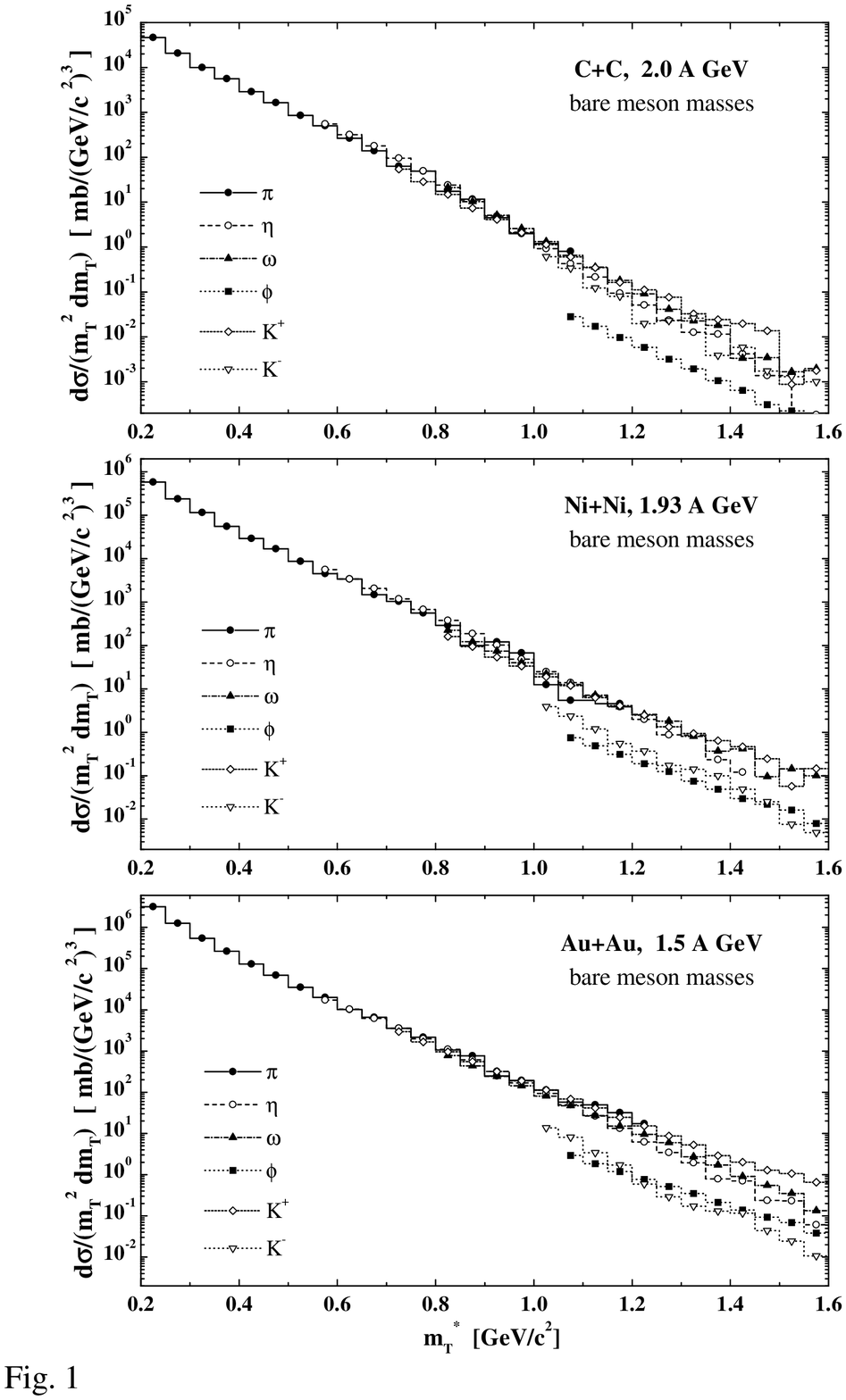,width=15cm,height=22cm}
\psfig{figure=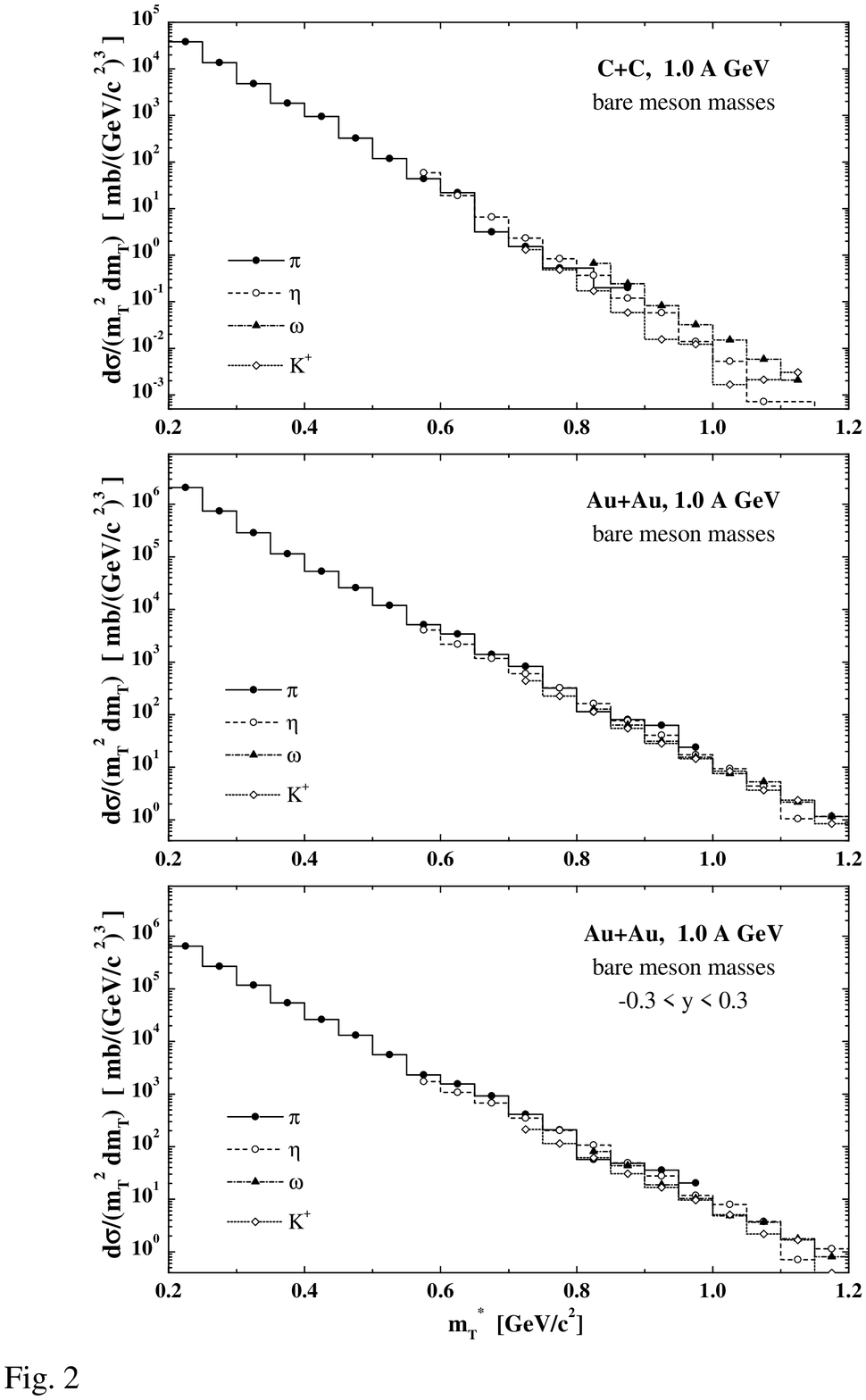,width=15cm,height=22cm}
\psfig{figure=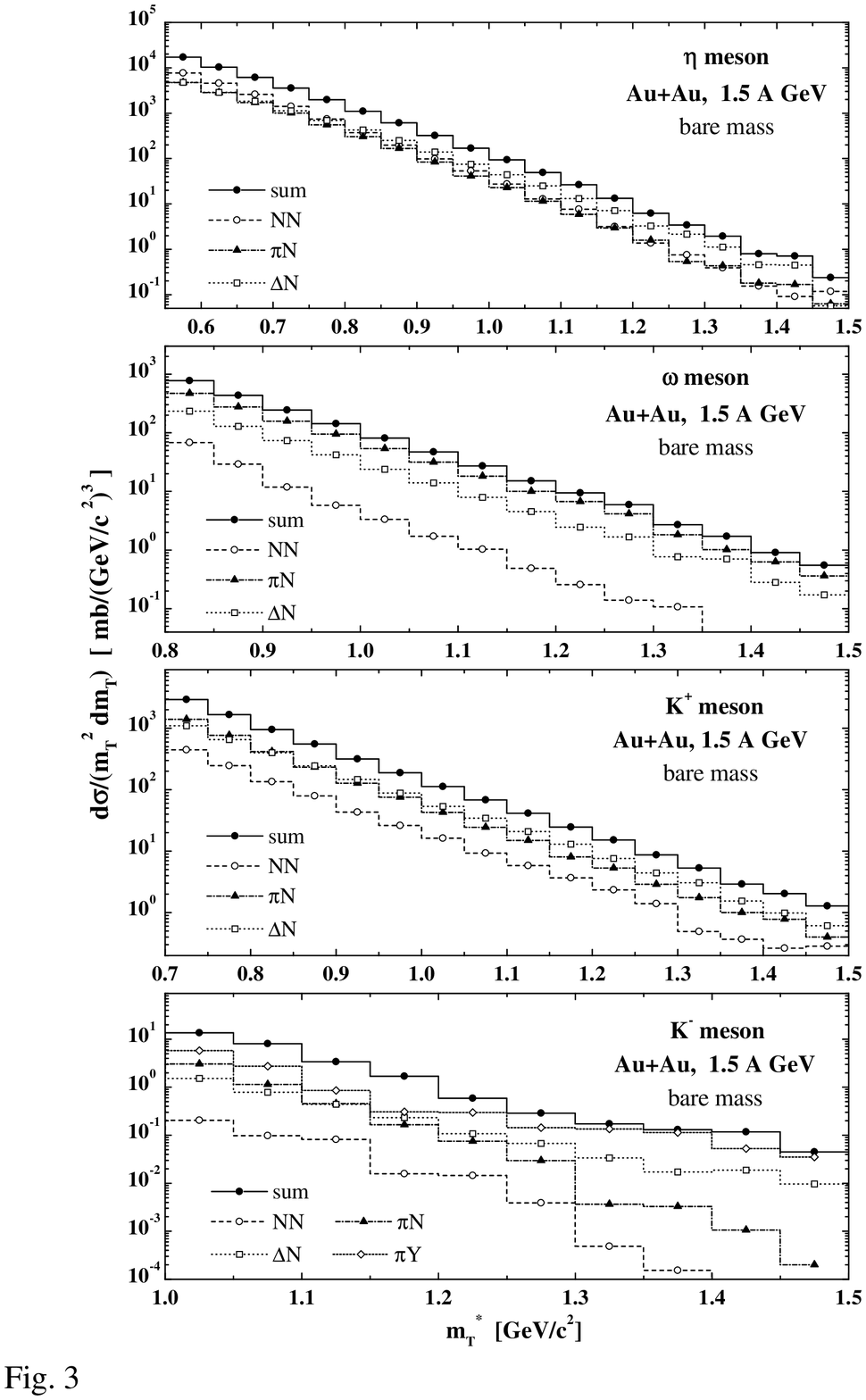,width=15cm,height=22cm}
\psfig{figure=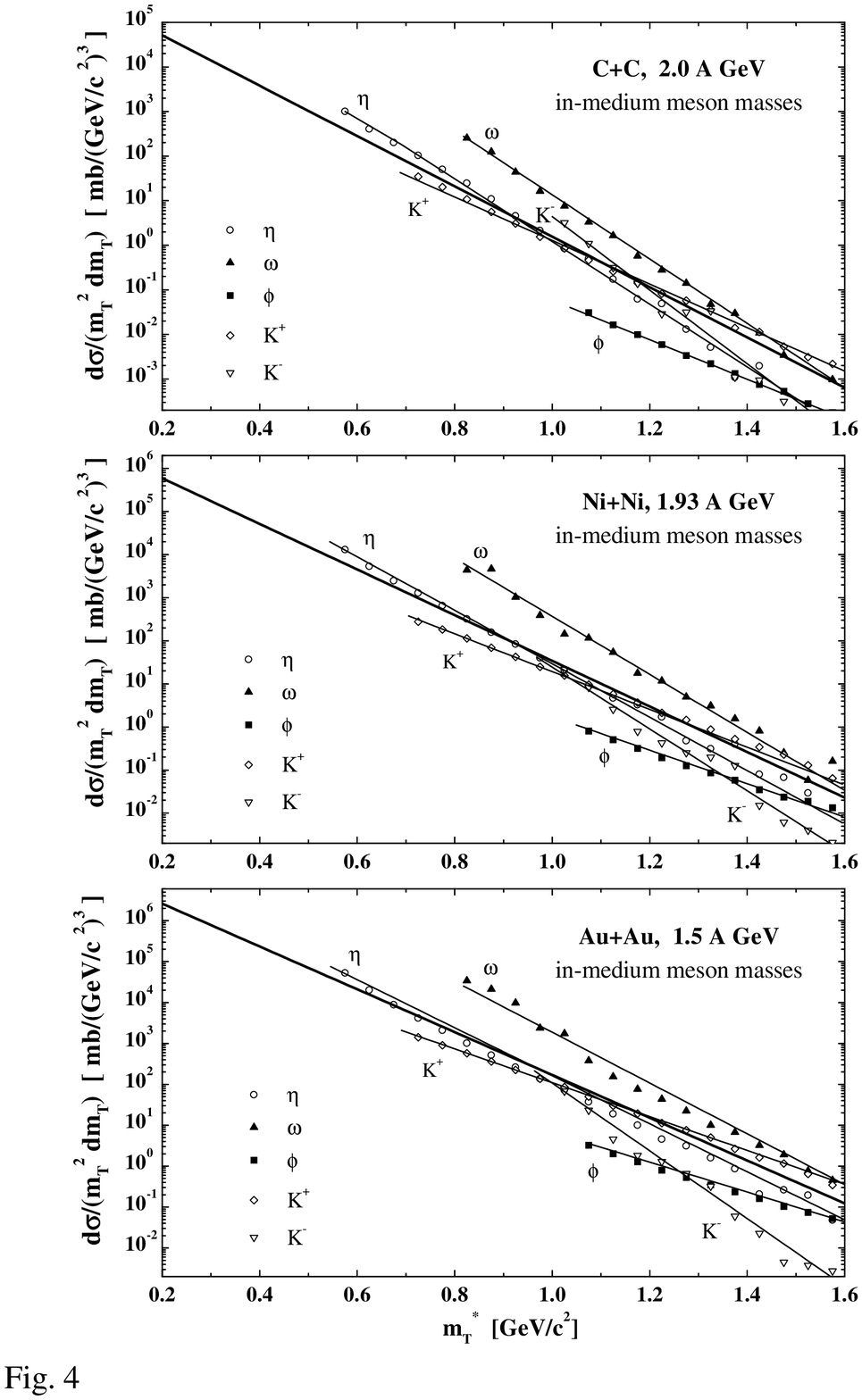,width=15cm,height=22cm}

\begin{thebibliography}{99}
\bibitem{H&L92}
	T. Hatsuda and S. Lee,  Phys. Rev. C 46 (1992) R34.
\bibitem{BrownRho}
	G.E. Brown and M. Rho, Phys. Rev. Lett. 66 (1991) 2720.
\bibitem{Klingl96}
	F. Klingl and W. Weise, Nucl. Phys. A 606 (1996) 329;
	F. Klingl, N. Kaiser and W. Weise, Nucl. Phys. A 624 (1997) 527.
\bibitem{Friman}
        B. Friman and H. J. Pirner, Nucl. Phys. A 617 (1997) 496.
\bibitem{RappNPA}
        R. Rapp, G. Chanfray and J. Wambach, Nucl. Phys. A 617 (1997) 472.
\bibitem{Peters}
        W. Peters, M. Post, H. Lenske, S. Leupold, and U. Mosel,
        nucl-th/9708004, Nucl. Phys. A, in press.
\bibitem{Kaplan}
	D.B. Kaplan and A.E. Nelson, Phys. Lett. B 175 (1986) 57.
\bibitem{Brown1}
	G.E. Brown, K. Kubodera, and M. Rho, Phys. Lett.   B 192 (1987) 273;
    	G.E. Brown, C.M. Ko, and K. Kubodera, Z. Phys. A 341 (1992) 301;
	G.E. Brown, C.-H. Lee, M. Rho and V. Thorsson, Nucl. Phys. A 567 (1994) 937.
\bibitem{Waas}
	T. Waas, N. Kaiser, and W. Weise,
	Phys. Lett. B 365 (1996) 12; B 379 (1996) 34.
\bibitem{CERES}
    	G. Agakichiev et al., Phys. Rev. Lett. 75 (1995) 1272.
\bibitem{Ullrich}
        Th. Ullrich et al.,  Nucl. Phys. A 610 (1996) 317c;
        A. Drees, Nucl. Phys. A 610 (1996) 536c.
\bibitem{HELIOS}
        M. A. Mazzoni, Nucl. Phys. A 566 (1994) 95c;
        M. Masera, Nucl. Phys. A 590 (1995) 93c;
        T. {\AA}kesson et al., Z. Phys. C 68 (1995) 47.
\bibitem{Li}
        G.Q. Li, C.M. Ko, and G.E. Brown, Phys. Rev. Lett. 75 (1995) 4007.
\bibitem{Cass95C}
        W. Cassing, W. Ehehalt, and C. M. Ko, Phys. Lett. B 363 (1995) 35;
        W. Cassing, W. Ehehalt, and I. Kralik,  Phys. Lett. B 377 (1996) 5.
\bibitem{Brat97}
        E.L. Bratkovskaya and W. Cassing, Nucl. Phys. A 619 (1997) 413.
\bibitem{QM96}
	"Quark Matter 96", Nucl. Phys. A 610 (1996) and Refs. therein.
\bibitem{CBRW97}
        W. Cassing, E.L. Bratkovskaya, R. Rapp, and J. Wambach,
        nucl-th/9708020, Phys. Rev. C, in press.
\bibitem{Schro}
	A. Schr\"oter et al., Z. Phys. A 350 (1994) 101.
\bibitem{KaoS}
	P. Senger and the KaoS Collaboration,
       Acta Physica Polonica  B 27 (1996) 2993;
	R.Barth et al., Phys. Rev. Lett. 78 (1997) 4007.
\bibitem{FOPI}
       J.L. Ritman et al., Z. Phys. A 352 (1995) 355;
	N. Herrmann, Nucl. Phys. A 610 (1996) 49c.
\bibitem{Kolix}
	G.Q. Li, C.M. Ko, and X.S. Fang, Phys. Lett. B 329 (1994) 149.
\bibitem{CBMTS97km}
	W. Cassing, E.L. Bratkovskaya, U. Mosel, S. Teis and A. Sibirtsev,
	Nucl. Phys.  A 614 (1997) 415.
\bibitem{LiLB97}
	G.Q. Li, C.-H. Lee, and G.E. Brown, nucl-th/9706057,
	Nucl. Phys. A, in press.
\bibitem{flow1}
	G.Q. Li, C.M. Ko and B.A. Li, Phys. Rev. Lett. 74 (1995) 235;
 	G.E. Brown, C.M. Ko and G.Q. Li,	nucl-th/9608039.
\bibitem{BCM97kp}
	E.L. Bratkovskaya, W. Cassing and U. Mosel,
	Nucl. Phys.  A 622 (1997) 593.
\bibitem{TAPSold}
        O. Schwalb et al., Phys. Lett. B 321 (1994) 20;
        F. D. Berg et al., Phys. Rev. Lett. 72 (1994) 977.
\bibitem{TAPS-CC}
        R. Averbeck et al., Z. Phys. A 359 (1997) 65.
\bibitem{TAPS-CaNi}
        M. Appenheimer et al., GSI Annual Report 1996, p.58.
\bibitem{Toneev}
        K. K. Gudima, M. Ploszajczak, V.D. Toneev,
	 Phys. Lett. B 328 (1994) 249.
\bibitem{BCRW97}
        E. L. Bratkovskaya, W. Cassing, R. Rapp, and J. Wambach,
        nucl-th/9710043.
\bibitem{Ehehalt}
        W. Ehehalt and W. Cassing, Nucl. Phys. A 602 (1996) 449.
\bibitem{Wolf90}
        Gy. Wolf, G. Batko, W. Cassing et al., Nucl. Phys. A 517 (1990) 615;
        Gy. Wolf, W. Cassing and U. Mosel, Nucl. Phys. A 552 (1993) 549.
\bibitem{Vetter}
        T. Vetter, A. Engel, T. Biro and U. Mosel,
	 Phys. Lett. B 263 (1991) 153.
\bibitem{WASA}
	 H. Cal\'en et al., Phys. Rev. Lett., in press;
	 S. H\"aggstr\"om, Ph.D. Thesis, Univ. of Uppsala,
	 Acta Universitatis Upseliensis 13, 1997.
\bibitem{SibCM97}
        A. Sibirtsev, W. Cassing and U. Mosel, Z. Phys. A 358 (1997) 357.
\bibitem{Sibirtsev}
	A. Sibirtsev and W. Cassing, nucl-th/9712009,
	Nucl. Phys. A (1997), in press.
\bibitem{WBCM97}
	Th. Weidmann, E.L. Bratkovskaya, W. Cassing and U. Mosel,
	nucl-th/9711004.
\bibitem{Cleyman}
	J. Clemans, D. Elliott, A. Ker\"anen, and E. Suhonen, nucl-th/9711066.
\bibitem{Stachel1}
	P. Braun-Munzinger et al., Phys. Lett. B 344 (1995) 43.
\bibitem{Stachel2}
	P. Braun-Munzinger et al., Phys. Lett. B 365 (1996) 1.
\bibitem{Metag}
	V. Metag, Prog. Part. Nucl. Phys. 30 (1993) 75;
	GSI-Preprint-97-43, Nucl. Phys. A, in press.
\bibitem{Teis}
	S. Teis, W. Cassing, M. Effenberger et al.,
  	Z. Phys. A 356 (1997) 421.
\bibitem{Pelte}
	D. Pelte et al., Z. Phys. A 359 (1997) 47.
\bibitem{Chung97}
	W.S. Chung, G.Q. Li, and C.M. Ko, nucl-th/9704002,
	Nucl. Phys. A, in press.
\end{thebibliography}
\end{document}